\documentclass[12pt,a4paper]{article}

\usepackage{amsmath}
\usepackage{amssymb}
\usepackage{tikz}
\usepackage{fixltx2e}
\usepackage{amsfonts}
\usepackage{graphicx}
\usepackage{subfig}
\usepackage{caption}
\usepackage{dcolumn}
\usepackage{bm}
\usepackage[T1]{fontenc}
\usepackage{textcomp}
\usepackage[
colorlinks,linkcolor=blue,citecolor=magenta]{hyperref}
\usepackage[top=3cm,right=2cm,bottom=2.5cm,left=2cm]{geometry}
\begin{document}
\title{ Hamiltonian formalism of the ghost free Tri(-Multi)gravity theory}

\author{Zahra Molaee$^{b}$\thanks{zmolaee@ipm.ir}, Ahmad Shirzad$^{a,b}$\thanks{shirzad@ipm.ir}\\
	{\it $^{a}$Department of Physics, Isfahan University of Technology, Iran}\\
{\it$^{b}$School of Particles and Accelerators,}\\
	{\it Institute for Research in Fundamental Sciences (IPM),}\\
	{\it P.O.Box 19395-5531, Tehran, Iran}}
\maketitle
\begin{abstract}
	We study the Hamiltonian structure of tri-gravity and four-gravity in the framework of ADM decomposition of the corresponding metrics. Hence we can deduce the general structure of the constraint system of multi-gravity. We will show it is possible and consistent to assume additional constraints which provide the needed first class constraints for generating  diffeomorphism  as well as enough second class constraints to omit the ghosts.

\end{abstract}
\section{Introduction}
There are some scenarios to insert mass inside the gravitational theory instead of considering dark matter which only interacts gravitationally. Massive gravity has a long history beginning with the pioneer paper of Fierz and Pauly. The main problem is  appearing the Boulware-Deser ghost \cite{BD} in most of primitive modifications of EH gravity. However, during the last decade there emerge hopes to have some ghost free massive models.

One of the well-known models in modified gravity is recognized as the dRGT (de Rham, Gabadadze and  Tolley) massive gravity \cite{drgt2}. It turns out that dRGT massive gravity based on a flat background metric is free of BD ghost in the non-linear level  \cite{Ha1}. This theory demonstrates  propagation of a  massive spin-2 field with  5 degrees of freedom.

In 2011, Hassan and Rosen presented the non-linear massive gravity based on a general background metric \cite{H1}. This model also includes  exactly five degrees of freedom. In fact, due to the special interaction term,  there exists additional constraints in the Hamiltonian analysis which eliminates the ghost mode. Using the full phase space description, which includes lapse and shift functions as physical variables, we have shown \cite{MS1} that indeed 10 second class constraints may emerge within the 20 dimensional phase space.

Soon after massive gravity, Hassan and Rosen introduced a bi-gravity model  by adding a dynamical term for the background metric to their massive gravity model \cite{Hassan1}. The Hamiltonian structure of HR bi-gravity was, however, a challenging issue. There was claims that consistency procedure of constraints leads to determining the lapse functions \cite{Kluson1}, \cite{Kluson2}, while we need enough first class constraints to generate the diffeomorphism symmetry. This
 issue was addressed  and resolved in refs \cite{HH2} and \cite{MS2}.

 In our recent work \cite{MS2} we showed that in the canonical investigation of the system, in the full 40 dimensional phase space, we may have two sets of four first class constraints as the generators of diffeomorphism, as well as two additional second class constraints which eliminate the ghost. However, we showed that at a critical point we meet a bifurcation problem where only one branch sounds physically acceptable. In other words, the theory by itself does not guide us naturally towards the physical branch.

Theories with more than two metrics are also of interest in the community \cite{s1}-\cite{s5}.
Theoretically, One may be interested to include several multiple spin-2 fields in her/his model, as is done for lower spins in field theory. In other words, one may ask why should we assume just  one massless spin-2 multiple; instead we can investigate possibility of several massive or massless multiples. On the other hand, there are also hopes for employing multi-gravity models in order to justify cosmological observations\cite{s6}-\cite{s13}. 

 For example one may generalize the HR bi-gravity to a multi-gravity theory with $\mathcal{N}  $ metrics, which involves, in the linear level, one massless spin-2 and $\mathcal{N}-1$ massive spin-2 gravitons. If there is no ghost, this theory would consist $5 \mathcal{N} -3$ dynamical degrees of freedom. This result, however, needs to be confirmed at the non linear level, too.
 
In ref \cite{sall} the authors  constructed  theories of multi-gravity in arbitrary space-time dimensions in vielbein formulation for  non-pairwise interactions.
It is now believed that
		the multi-gravity theories introduced in \cite{sall} are in general not ghost free, as argued in \cite{de1}. An exception is theories where the  metrics interact only pairwise and those pairwise interactions are
		bi-metric interactions. These are the type of models considered in the  present paper (the models considered in \cite{sall} have, generically, more  than pairwise interactions, with each interaction term containing the
		product of more than two vielbeins). An exception to the general argument of \cite{de1} was found in ref \cite{s3} which obtained specific non-pairwise multi-gravity interactions that are ghost free. In
		this case, the absence of ghost was argued based on a Hamiltonian analysis in a 3+1 decomposition.  
	   However, a perfect Hamiltonian analysis of tri-gravity and multi-gravity, based on ADM decomposition of metric variables, has not been performed yet. This is our aim in this paper, where we try to generalize our investigations concerning bi-gravity \cite{MS2} first to three and four gravity and then induce the results for multi-gravity.  In order to have a complete dynamical description, we have performed our investigations in the full phase space which consists $20\mathcal{N} $ phase space variables including the lapse and shift functions as physical variables.  Our main result is that we may have enough first class constraints to generate the required guage symmetry as well as additional constraints to omit the ghost, but this may happen in a special branch of the bifurcation point, i.e. the physical branch. In the following three sections we will study three, four and multi gravities respectively.

\section{ The  HR  tri-gravity theory}
The tri-gravity, as the first extension of bi-gravity, is described by the following action \cite{Hassan1}
\begin{eqnarray}&&
S=M^{2}_{g} \int d^{4}x \sqrt{-g} 
\mathcal{R}(g)+M^{2}_{f}\int d^{4}x \sqrt{-f} 
\mathcal{R}(f)+M^{2}_{h} \int d^{4}x \sqrt{-h} 
\mathcal{R}(h) \nonumber\\ && 
+2m^{4} \int d^{4}x \sqrt{-g}  \sum_{n=0}^{4} \beta^{g}_{n}e_{n}(\mathcal{X})+2m^{4} \int d^{4}x \sqrt{-f}  \sum_{n=0}^{4} \beta^{f}_{n}e_{n}(\mathcal{Y}). \label{a1}
\end{eqnarray}
where $\beta^{g}_{n}$ and $ \beta^{f}_{n} $ are free parameters, $m$ is a mass parameter and $M_{g},M_{h}$ and $ M_{f}$ are three different Plank masses. The matrices $\mathcal{X} $ and $\mathcal{Y} $ are define as $\mathcal{X} \equiv \sqrt{(g^{-1}h)^{\mu}_{\nu}}$ and  $\mathcal{Y} \equiv \sqrt{(f^{-1}h)^{\mu}_{\nu}}$.
The elementary symmetric polynomials $e_{n}(\mathcal{X})$  are shown as follow
 \begin{eqnarray}&&
e_{0}(\mathcal{X})=1,\label{a2}
\nonumber\\ &&
e_{1}(\mathcal{X})=[\mathcal{X}],\label{a3}
\nonumber\\ &&
e_{2}(\mathcal{X})=\frac{1}{2}([\mathcal{X}]^{2}-[\mathcal{X}^{2}]),\label{a4}
\nonumber\\ &&
e_{3}(\mathcal{X})=\frac{1}{6}([\mathcal{X}]^{3}-3[\mathcal{X}][\mathcal{X}^{2}]+2[\mathcal{X}^{3}]),\label{a5}
\nonumber\\ &&
e_{4}(\mathcal{X})=\frac{1}{24} ( [\mathcal{X}]^{4}-6[\mathcal{X}]^{2}[\mathcal{X}^{2}] + 
3[\mathcal{X}^{2}]^{2}+8[\mathcal{X}][\mathcal{X}^{3}]-6[\mathcal{X}^{4}]),\label{a6}
\nonumber\\ &&
e_{i}(\mathcal{X})=0,\ \ i > 4,\label{a7}
\end{eqnarray}
where $[\mathcal{X}] \equiv Tr(\mathcal{X})$ and so on.
In ADM decomposition, the first metric and its inverse   have the following form \cite{pad}, 
\begin{equation}
g_{\mu\nu}=
\left( \begin{array}{cr}
-N^{2}+N_{i}N^{i} & N_{i}  \\
N_{i} &  g_{ij} \\
\end{array}\right),
g^{\mu\nu}=
\left( \begin{array}{cr}
-N^{-2}       & N^{i}N^{-2}  \\   
N^{j}N^{-2}      & g^{ij}-N^{i}N^{j}N^{-2}  \\
\end{array}\right),\label{bg}
\end{equation}
where $ N, N^{i} $ are lapse and shift functions respectively. We assume the similar form for $f^{\mu\nu} $ and  $h^{\mu\nu} $ with $ M,M^{i}$ and $L, L^{i} $  as lapses and shifts.

In this paper we study only minimal model of the interaction term where  $\beta_i=0$ for $i\ne 1$ and $\beta_1=1$. Let us use variables ($ N, n^{i} $) instead of ($ N, N^{i} $) \cite{Hassan1} as
\begin{equation}
N^{i}=L n^{i}+L^{i}+N D_{j}^{i}n^{j},\label{a15}
\end{equation}
where the $3\times3$ matrix $D^{i}_{\ j}$ has the following form 
\begin{equation}
 D^{i}_{\ j}=\sqrt{g^{id}h_{dm}W^{m}_{n}}(W^{-1})^{n}_{j},\label{a18}\hspace{5mm}
W^{l}_{\ j}=[1-n^{k}h_{km}n^{m}]\delta^{l}_{j}+n^{l}h_{mj}n^{m}.
\end{equation}
However, one may wonder if the (3+1) decomposition used in Eq. (\ref{bg}) is valid simultaneously for all three metrics. Also one should be convinced about existing the matrix square roots in Eq. (\ref{a18}). As discussed in ref. \cite{kocic}, the condition is overlapping of the null cones of the metrics. In other words, the $ g-h $ and $ f-h $ cones should have non-zero intersections. 
Here, we assume such conditions are satisfied and the action (\ref{a1}) and decomposition (\ref{bg})  are meaningful.

 It turns out that the first interaction term containing $ \sqrt{g^{-1}h} $ would be linear in $N$. In the same way, using variables ($ M,m^{i} $) as
\begin{equation}
 M^{i}=L m^{i}+L^{i}+M \tilde{D}_{j}^{i}m^{j},\label{a25}
\end{equation}
makes the  whole action linear with respect to the lapses  $N$, $ M$ , $L$ and shifts  $L^{i}$. 
 The expression of  $\tilde{D}^{i}_{\ j}$ in Eq. (\ref{a25}) is similar to Eq. (\ref{a15}) where $g_{ij}$ is replaced by $f_{ij}$. 
Thus, the Lagrangian density reads
\begin{eqnarray}&&
\mathcal{L}=M^{2}_{g}\pi^{ij}\partial_{t}g_{ij}+M^{2}_{f}p^{ij}\partial_{t}f_{ij} +M^{2}_{h}\rho^{ij}\partial_{t}h_{ij}-(L^{i} \mathcal{R}_{i} +N \phi_{1}+ M \phi_{2} +L\phi). \label{bi4}
\end{eqnarray} 
in which
\begin{eqnarray}&&
\phi_{1}=M^{2}_{g}\mathcal{R}_{0}^{g}+M^{2}_{g} D^{i}_{\ k} n^{k} \mathcal{R}_{i}^{g}+2m^{4}(\sqrt{g}\sqrt{x} D_{\ k}^{ k}),\label{b557}\\ &&
\phi_{2}=M^{2}_{f}\mathcal{R}_{0}^{f}+ M^{2}_{h} m^{j}\tilde{D}^{i}_{\ j}\mathcal{R}_{i}^{f}+2m^{4}(\sqrt{f}\sqrt{y}\tilde{D}_{\ k}^{ k}),\\ && \label{bii7}
\phi=M^{2}_{f}m^{i}\mathcal{R}_{i}^{f}+ M^{2}_{g} n^{i}\mathcal{R}_{i}^{g}+M^{2}_{h}\mathcal{R}_{0}^{h}+2m^{4}\sqrt{f}\sqrt{y}+2m^{4}\sqrt{g}\sqrt{x},\label{bi6} \\ &&
\mathcal{R}_{i}= M^{2}_{g} \mathcal{R}_{i}^{g}+M^{2}_{f}\mathcal{R}_{i}^{f}+ M^{2}_{h} \mathcal{R}_{i}^{h},\label{bii16}
\end{eqnarray}
where $ x=1-n^{i}h_{ij}n^{j}$ and  $ y=1-m^{i}h_{ij}m^{j}$. 
 The momentum fields  conjugate to $ g_{\mu\nu} $, $ f_{\mu\nu} $ and $ h_{\mu\nu} $ are respectively  
\begin{eqnarray}&&
\pi^{ij}=-\sqrt{g}(\mathcal{U}^{ij}-g^{ij}\mathcal{U})\label{c1},\label{bi3}\\&&
p^{ij}=-\sqrt{f}(\mathcal{V}^{ij}-f^{ij}\mathcal{V}),\label{bi1}\\&&
\rho^{ij}=-\sqrt{h}(\mathcal{W}^{ij}-h^{ij}\mathcal{W}).\label{bi8}
\end{eqnarray}
where $\mathcal{U}^{ij}$ , $\mathcal{V}^{ij}$ and $ \mathcal{W}^{ij} $ are extrinsic curvatures related to the metrics $ g, f$ and $ h $. The velocities are present in the first two terms of the Eqs. (\ref{b557}-\ref{bi6})
due to the Hilbert-Einstein terms  of the action.
Obviously,  we have 12 primary constraints as
\begin{eqnarray}&&
P_{L_{i}}\approx 0, P_{M}\approx 0,P_{N} \approx 0,P_{L}\approx 0,P_{n^{i}}\approx 0,P_{m^{i}}\approx 0.\label{bi}
\end{eqnarray}
Since, the interaction terms do not involve velocities,  using Eqs. (\ref{bi3}-\ref{bi8}) expressions $\mathcal{R}^{(g)}_{0}$ , $\mathcal{R}^{(g)}_{i}$,  for instant, can be written in terms of  momentum fields as
\begin{equation}
\mathcal{R}_{0}^{(g)}= M_{g}^{2}\sqrt{g}\mathcal{R}+\dfrac{1}{ M_{g}^{2}\sqrt{g}}(\frac{1}{2}\pi^{2}-\pi^{ij}\pi_{ij}),\hspace{10mm} \mathcal{R}_{i}^{(g)}=2\sqrt{g}g_{ij} \triangledown _{k}(\frac{\pi^{jk}}{\sqrt{g}}),\label{a14}
\end{equation}
with similar expressions for $\mathcal{R}^{(f)}_{0}$, $\mathcal{R}^{(f)}_{i}$ and $\mathcal{R}^{(h)}_{i}$,$\mathcal{R}^{(h)}_{0}$  in terms of the $f$ and $ h $-metrics. Note also that  $ \sqrt{-^{(4)}g} =N\sqrt{g} $, where $ g\equiv det(g_{ij}) $. 
Hence, the total Hamiltonian is 
\begin{eqnarray} &&
\mathcal{H}_{T}=\mathcal{H}_{c}+wP_{L}+uP_{N}+vP_{M}+w^{i}P_{L^{i}}+u^{i}P_{n^{i}}+v^{i}P_{m^{i}}, \label{k111}
\end{eqnarray} 
where $ u,v,u_{i},w,w^{i} $ and $v_{i}$ are 12 undetermined Lagrange multipliers (12 fields, in fact) and the canonical Hamiltonian  reads
\begin{equation}
\mathcal{H}_{c}=L^{i} \mathcal{R}_{i} + M \phi_{2}  + N \phi_{1}+L \phi.\label{bi5}
\end{equation}

Since $N$, $M$, $L$ and $L_{i}$ appear linearly  in the canonical Hamiltonian, consistency of the primary constraints $P_{M}$, $P_{N}$,$P_{L}$ $P_{L_{i}}$  gives 6 secondary constraints as follows
\begin{eqnarray} &&
\lbrace P_{N},\mathcal{H}_{T} \rbrace= -\phi_{1}\approx 0,
\\&& \lbrace P_{M},\mathcal{H}_{T} \rbrace=-\phi_{2} \approx 0,\\&&
\lbrace P_{L},\mathcal{H}_{T} \rbrace= - \phi\approx 0,
\\&& \lbrace  P_{L^{i}},\mathcal{H}_{T} \rbrace=-\mathcal{R}_{i}\approx 0,
\end{eqnarray}
 while, for consistency of $P_{n^{i}}$ and $P_{m^{i}}$, we derive directly 
\begin{eqnarray} &&
\lbrace P_{n^{i}},\mathcal{H}_{c}\rbrace \equiv -\left( L \delta^{i}_{k}+N\frac{\partial (D^{i}_{j}n^{j})}{\partial n^{k}}\right)\mathcal{S}^{(1)}_{i}  \approx 0,\label{ti3}
\end{eqnarray}
and
\begin{eqnarray} &&
\lbrace P_{m^{i}},\mathcal{H}_{c}\rbrace \equiv -\left( L \delta^{i}_{k}+M\frac{\partial (\tilde{D}^{i}_{j}m^{j})}{\partial m^{k}}\right)\mathcal{S}^{(2)}_{i}  \approx 0.\label{ti131}
\end{eqnarray}
Eqs. (\ref{ti3}) and (\ref{ti131})  lead to the secondary constraints
\begin{eqnarray} &&
\mathcal{S}^{(1)}_{i}=M^{2}_{g}\mathcal{R}_{k}(g)-2m^{4}\sqrt{g}n^{l}h_{lj}\delta^{j}_{\ k}x^{-1/2}\approx 0,\label{ti11}
\end{eqnarray}
and
\begin{eqnarray} &&
\mathcal{S}^{(2)}_{i}=M^{2}_{f}\mathcal{R}_{k}(f)-2m^{4}\sqrt{f}m^{l}h_{lj}\delta^{j}_{\ k}y^{-1/2}\approx 0.\label{ti31}
\end{eqnarray}
The matrices in the parenthesis of  Eqs. (\ref{ti3}) and (\ref{ti131}) are the Jacobian of the transformation given in Eqs. (\ref{a15}) and (\ref{a25}) that are invertible. We can determine $ n_{i} $ and $ m_{i} $ respectively from $\mathcal{S}^{(1)}_{i}=0$ and $\mathcal{S}^{(2)}_{i} =0 $. Thus, $ \mathcal{S}^{(1)}_{i},\mathcal{S}^{(2)}_{i}, P_{n^{i}}, P_{m^{i}} $ are 12 second class constraints. 

We should now investigate the consistency of the secondary constraints $  \phi_{1}, \phi_{2} ,  \phi $ and $ \mathcal{R}_i $. Let's begin with consistency of $ \mathcal{R}_i $. Direct calculation shows that
\begin{eqnarray} &&
\lbrace \mathcal{R}_i,\mathcal{H}_T\rbrace 
\approx 0.
\end{eqnarray}
Since $ \mathcal{R}_i $'s are the sum of momentum constraints of individual Hilbert-Einstein terms, they are a set of first class constraints which, together with the primary constraints $ P_{L^i}$, generate the spatial diffeomorphisms. Similar to bi-gravity model, here we can show directly \cite{Hassan1}
\begin{eqnarray} &&
\lbrace \phi_{1},\phi_{1}\rbrace 
 \approx \lbrace \phi_{2} ,\phi_{2} \rbrace 
 \approx \lbrace  \phi, \phi \rbrace 
 \approx 0.
\end{eqnarray}
As we will see below, these relations are crucial in omitting the ghosts. Direct calculation shows that 
\begin{eqnarray} &&
\lbrace \phi_{1},\phi_{2} \rbrace = 0
,  \hspace{5mm}
\lbrace  \phi_{1}, \phi \rbrace \equiv \psi_{1}
, \hspace{5mm}
\lbrace \phi_{2} , \phi\rbrace \equiv \psi_{2},
\end{eqnarray}
in which the exact form of $\psi_{1}  $ and $ \psi_{2} $ are as follows 
\begin{eqnarray} &&
\psi_{1}=\frac{m^4}{M^2_g}(g_{mn}\pi-2\pi_{mn})U^{mn}+2m^4 \sqrt{g}g_{ni}D^{i}_{k}n^{k}\triangledown_{m}U^{mn}+\nonumber\\&&
(\mathcal{R}_{j}(g)D^{i}_{k}n^{k}-2m^{4}\sqrt{g}g_{ik}\bar{V}^{ki})\triangledown_{i}n^{j}+\sqrt{g}(\triangledown_{i}(\mathcal{R}^{0}(g)/\sqrt{g})+\triangledown_{i}(\mathcal{R}_{j}(g)/\sqrt{g})D^{j}_{k}n^{k})n^{i}\nonumber\\&&-\frac{m^{4}}{M^{2}_{h}}\frac{\sqrt{g}}{\sqrt{h}}(h_{mn}\rho-2\rho_{mn})\tilde{F}^{mn},\\&&
\psi_{2}=\frac{m^4}{M^2_f}(f_{mn}p-2p_{mn})\tilde{U}^{mn}+2m^4 \sqrt{f}f_{ni}\tilde{D}^{i}_{k}m^{k}\tilde{\triangledown}_{m}\tilde{U}^{mn}+\nonumber\\&&
(\mathcal{R}_{j}(f)\tilde{D}^{i}_{k}m^{k}-2m^{4}\sqrt{f}f_{ik}\tilde{V}^{ki})\tilde{\triangledown}_{i}m^{j}+\sqrt{f}(\tilde{\triangledown}_{i}(\mathcal{R}^{0}(f)/\sqrt{f})+\tilde{\triangledown}_{i}(\mathcal{R}_{j}(f)/\sqrt{f})\tilde{D}^{j}_{k}m^{k})m^{i}\nonumber\\&&-\frac{m^{4}}{M^{2}_{h}}\frac{\sqrt{f}}{\sqrt{h}}(h_{mn}\rho-2\rho_{mn})\bar{F}^{mn}, \label{psi}
\end{eqnarray}
where
\begin{eqnarray} &&
U^{mn}=-\sqrt{x}g^{mn}, \hspace{10mm} \tilde{U}^{mn}=-\sqrt{y}f^{mn},\nonumber\\&&
\bar{V}^{ki}=g^{kj}(-\frac{h_{jl}}{\sqrt{x}}(D^{-1})^{l}_{r}g^{ri}),\nonumber\\&&
\tilde{F}^{mn}=-\frac{(D^{-1})^{m}_{i}g^{ni}-n^{i}n^{m}D^{n}_{i}}{\sqrt{x}},\nonumber\\&&
\tilde{V}^{ki}=f^{kj}(-\frac{h_{jl}}{\sqrt{y}}(\tilde{D}^{-1})^{l}_{r}f^{ri}),\nonumber\\&&
\bar{F}^{mn}=-\frac{(\tilde{D}^{-1})^{m}_{i}f^{ni}-m^{i}m^{m}\tilde{D}^{n}_{i}}{\sqrt{y}}.
\end{eqnarray}
Hence, consistency of the secondary constraints $ \phi_{1}, \phi_{2}  $ and $  \phi $  leads to the following set of equations
\begin{equation}
\left( \begin{array}{ccc}
0      & 0 & \psi_{1}\\   
  0   & 0 & \psi_{2}  \\
   -\psi_{1} & -\psi_{2} & 0 \\
\end{array}\right)
\left( \begin{array}{c}
N     \\   
M  \\
L \\
\end{array}\right)=0, \label{gf}
\end{equation}
In the special case of tri-gravity,  rank of the matrix on the l.h.s. of Eqs. (\ref{gf}) may not exceed two, regardless of the form of  $\psi_{1}$ and $\psi_{2}$.

If $ \psi_{1}$ and/or $ \psi_{2} $ does not vanish, then at least the lapse function $L$ should vanish. This is not valid physically since the volume element vanishes and the inverse metric would be ill defined. Hence, the only physically acceptable possibility is that  $ \psi_{1}\approx 0$ and $ \psi_{2} \approx 0 $. In this case Eq. (\ref{gf}) holds identically. 

 However, we need to consider consistency of the new constraints $\psi_{1}$ and $\psi_{2}$. To this end we have 
\begin{eqnarray} &&
\left( \begin{array}{c}
\dot{\psi_{1}}     \\   
\dot{\psi_{2}}  \\
\end{array}\right)=
\left( \begin{array}{cc}
\mathcal{M}_{11}    & \mathcal{M}_{12}   \\   
\mathcal{M}_{21}        & \mathcal{M}_{22}    \\
\end{array}\right)
\left( \begin{array}{c}
N     \\   
M  \\
\end{array}\right)+\left( \begin{array}{c}
\mathcal{M}_{1L}     \\   
\mathcal{M}_{2L}  \\
\end{array}\right)L=0, \label{m1}
\end{eqnarray}
where $ \mathcal{M}_{11}=\{\psi_{1},\phi_{1}\} $, $\mathcal{M}_{12}=\{\psi_{1},\phi_{2} \}  $,  $\mathcal{M}_{1L} = \{\psi_{1},  \phi\} $,  $ \mathcal{M}_{21}=\{\psi_{2},\phi_{1}\} $, $\mathcal{M}_{22}=\{\psi_{2},\phi_{2} \}  $ and $\mathcal{M}_{2L} = \{\psi_{2},  \phi\} $. 
These are two equations for three unknowns $ M $, $ N $ and $ L $. Considering $ L $ as the arbitrary lapse function, we can determine $ N $ and $ M $ in terms of $ L $ as follows 
\begin{eqnarray} &&
\left( \begin{array}{c}
N     \\   
M  \\
\end{array}\right)=-\left( \begin{array}{cc}
\mathcal{M}_{11}    & \mathcal{M}_{12}   \\   
\mathcal{M}_{21}        & \mathcal{M}_{22}    \\
\end{array}\right)^{-1}
\left( \begin{array}{c}
\mathcal{M}_{1L}     \\   
\mathcal{M}_{2L}  \\
\end{array}\right)L.
\end{eqnarray}
Let us change the lapse functions from the very beginning as $ \bar{N} $, $ \bar{M} $ and $ L $ such that 
\begin{eqnarray} &&
\left( \begin{array}{c}
\bar{N}     \\   
\bar{M}  \\
\end{array}\right)=\left( \begin{array}{c}
N     \\   
M  \\
\end{array}\right)+\left( \begin{array}{cc}
\mathcal{M}_{11}    & \mathcal{M}_{12}   \\   
\mathcal{M}_{21}        & \mathcal{M}_{22}    \\
\end{array}\right)^{-1}
\left( \begin{array}{c}
\mathcal{M}_{1L}     \\   
\mathcal{M}_{2L}  \\
\end{array}\right)L.\label{jh}
\end{eqnarray}
At the same time let us introduce a new combination of second level constraints as 
\begin{eqnarray} &&
 \phi^{\prime}= \phi-\left( \begin{array}{c}
\mathcal{M}_{1L} ,\  \mathcal{M}_{2L}   \\
\end{array}\right)\left( \begin{array}{cc}
\mathcal{M}_{11}    & \mathcal{M}_{12}   \\   
\mathcal{M}_{21}        & \mathcal{M}_{22}    \\
\end{array}\right)^{-1^{T}}\left( \begin{array}{c}
\phi_{1}    \\   
\phi_{2}  \\
\end{array}\right).
\end{eqnarray}
Hence the canonical Hamiltonian reads 
\begin{eqnarray} &&
\mathcal{H}_{c}=\bar{N}\phi_{1}+\bar{M}\phi_{2} +L \phi^{\prime}+L^{i} \mathcal{R}_{i}.
\end{eqnarray}
Now the second level constraints due to consistency of $ P_{\bar{N}} $, $ P_{\bar{M}} $, $ P_{L} $ read $ \phi_{1} $, $ \phi_{2}  $ and $  \phi^{\prime} $ respectively. Direct calculation shows that $  \phi^{\prime} $ commutes with all the existing constraints $ \phi_{1} $, $ \phi_{2}  $, $ \psi_{1} $ and $ \psi_{2} $. Hence, the set $ P_{L} $ and $  \phi^{\prime} $ are our desired first class constraints.

 Consistency of the remaining constraints $\psi_{1} $ and $ \psi_{2}  $ gives $ \bar{M} \{\psi_{1},\phi_{2} \} +\bar{N}\{\psi_{1},\phi_{1}\}\approx 0$ and $\bar{N} \{\psi_{2},\phi_{1}\}+\bar{M}\{\psi_{2},\phi_{2} \}\approx 0$ respectively. In this way consistency of the second level constraint as well as the new constraints are satisfied by assuming the last level constraints $ \bar{N}=\bar{M}\approx 0 $, which determine the Lagrange multipliers under their own consistency conditions.
 Note from Eq. (\ref{jh}) that none of the lapse functions do vanish during the consistency procedure. In fact, $ \bar{N}\approx 0 $ and $ \bar{M}\approx 0 $ means that $ M $ and $ N $ depend on $ L $, which is a gauge parameter by itself. 
 
  Hence, we find all together two first class constraints $ P_{L} $ and $  \phi' $ plus eight second level constraints $ P_{\bar{N}} $,$ P_{\bar{M}} $, $ \phi_{1}$, $ \phi_{2} $, $ \psi_{1} $, $ \psi_{2}$, $ \bar{N} $ and $ \bar{M} $. Taking into account 6 first class constraints $ P_{L^{i}} $ and $ \mathcal{R}_{i} $ and 12 second class constraints $ P_{n^{i}} $, $ P_{m^{i}} $, $ \mathcal{S}^{(2)}_{i} $ and $ \mathcal{S}^{(1)}_{i} $, found previously,  we have finally $ 8 $ first class and 20 second class constraints. Using the master formula \cite{zms2}
\begin{eqnarray} &&
\# DOF=N-2 \times First\ Class -Second \ Class,\label{mf}
\end{eqnarray}
in which $ N $ is total number of phase space variables,
 the number of degrees of freedom is
\begin{eqnarray} &&
\# DOF=60-2 \times 8-20=24,
\end{eqnarray}
which corresponds to  $ 12 $ degrees of freedom in configuration space. These are related to  two massive and one massless graviton. In this way every thing is satisfactory and we have both diffeomorphism symmetry generators and appropriate number of dynamical variables.
As is seen, the original lapse function fortunately do not vanish; hence non of the metrics come out to be singular.

 \section{ The  HR  four-gravity theory}
In this section we consider the Hamiltonian structure of the four-gravity. As we will see the results are similar to tri-gravity model.  The interaction terms of the four-gravity action can be written as 
\begin{eqnarray}&&
2m^{4}\left(  \sqrt{-g}  \sum_{n=0}^{4} \beta^{g}_{n}e_{n}(\mathcal{X})+  \sqrt{-f}  \sum_{n=0}^{4} \beta^{f}_{n}e_{n}(\mathcal{Y})+ \sqrt{-l}  \sum_{n=0}^{4} \beta^{l}_{n}e_{n}(\mathcal{Z})\right). \label{a11}
\end{eqnarray}
where $\beta^{g}_{n}$, $\beta^{l}_{n}$ and $ \beta^{f}_{n} $ are free parameters, $m$ is a mass parameter and the matrices $\mathcal{X}$, $\mathcal{Y}$ and $\mathcal{Z} $  are $ \sqrt{(g^{-1}h)^{\mu}_{\nu}}$,  $ \sqrt{(f^{-1}h)^{\mu}_{\nu}}$ and $ \sqrt{(l^{-1}h)^{\mu}_{\nu}}$ respectively.

Similar to the previous section, we only consider the minimal model of the interaction terms where  $\beta^{(i)}_{1}=1$.
Let us consider the redefinitions $N^{i}, M^{i}$ as given in Eqs. (\ref{a15}, \ref{a25}) and $‌Q^{i}$ as  
\begin{equation}
 Q^{i}=L q^{i}+L^{i}+Q \tilde{\tilde{D}}_{j}^{i}q^{j},
\end{equation}
where $ Q $ and $ Q^{i} $ are lapse and shift functions for forth metric and  $ \tilde{\tilde{D}} $ is similar to Eq. (\ref{a18}) in which $ g $ is replaced with $ l $.
The variables $N$, $M$, $L,‌ Q$ and $L_{i}$ appear linearly  in the canonical Hamiltonian, i.e.
\begin{equation}
\mathcal{H}_{c}=L^{i} \mathcal{R}_{i} + M \phi_{2}+ N \phi_{1}+Q \phi_{3}+L\phi,\label{bi55}
\end{equation}
where
\begin{eqnarray}&&
 \phi_{1}=M^{2}_{g}\mathcal{R}_{0}^{g}+M^{2}_{g} D^{i}_{\ k} n^{k} \mathcal{R}_{i}^{g}+2m^{4}(\sqrt{g}\sqrt{x} D),\\ &&\label{bi57}
 \phi_{2}=M^{2}_{f}\mathcal{R}_{0}^{f}+ M^{2}_{h} m^{j}\tilde{D}^{i}_{\ j}\mathcal{R}_{i}^{f}+2m^{4}(\sqrt{f}\sqrt{y}\tilde{D}),\\ && \label{bi7}
  \phi_{3}=M^{2}_{l}\mathcal{R}_{0}^{l}+M^{2}_{l}  \tilde{ \tilde{D}}^{i}_{\ k} q^{k} \mathcal{R}_{i}^{l}+2m^{4}(\sqrt{l}\sqrt{z}   \tilde{\tilde{D}}),\\ &&
  \phi=M^{2}_{f}m^{i}\mathcal{R}_{i}^{f}+M^{2}_{l}q^{i}\mathcal{R}_{i}^{l}+ M^{2}_{g} n^{i}\mathcal{R}_{i}^{g}+ M^{2}_{h}\mathcal{R}_{0}^{h}\hspace{5mm}\nonumber\\ &&\hspace{6mm}+2m^{4}\sqrt{f}\sqrt{y}+2m^{4}\sqrt{g}\sqrt{x}+2m^{4}\sqrt{l}\sqrt{z},\\ &&
\mathcal{R}_{i}= M^{2}_{g} \mathcal{R}_{i}^{g}+M^{2}_{f}\mathcal{R}_{i}^{f}+ M^{2}_{h} \mathcal{R}_{i}^{h}+ M^{2}_{l} \mathcal{R}_{i}^{l},
\label{bii6}
\end{eqnarray}
in which $ x $ and $ y $ are given as before  and $ z=1-q^{i}l_{ij}q^{j} $ for the last metric. 
The total Hamiltonian reads 
\begin{eqnarray} &&
\mathcal{H}_{T}=\mathcal{H}_{c}+r_{L}P_{L}+u_{N}P_{N}+v_{M}P_{M}+w_{Q}P_{Q}+r^{i}P_{L^{i}}+u^{i}P_{n^{i}}+v^{i}P_{m^{i}}+w^{i}P_{q^{i}}, \label{k4}
\end{eqnarray} 
where $ u_{N},v_{M},u^{i},r_{L},r^{i},w_{Q},w^{i} $ and $v^{i}$ are 16 undetermined Lagrangian multipliers (16 fields, in fact). 
Consistency of the primary constraints $P_{M}$, $P_{N}$, $P_{L}$, $P_{Q}$ and $P_{L_{i}}$  gives 7 secondary constraints as  $ \phi_{3}$, $ \phi$, $ \phi_{2}$, $\phi_{1} $ and  $\mathcal{R}_{i}$.
Similar as tri-gravity one can see $\{\mathcal{R}_{i}, \mathcal{H}_{T} \}\approx 0$, which shows that $\mathcal{R}_{i}  $ and $P_{L^{i}}  $ are first class constraints which generate spatial diffeomorphism. Moreover, it is directly seen that
$  \mathcal{S}_{(k)}^{i} $ and $P_{n_{(k)}^{i}}  $  are second class constraints  in which $n_{(k)}^{i}$ denote $ n^{i} $, $ m^{i} $ and $ q^{i} $ respectively.
 
In order to derive the important part of the  consistency of the secondary constraints, first note that as before
\begin{equation}
\lbrace  \phi_{1},  \phi_{1}\rbrace \approx \lbrace  \phi_{2},  \phi_{2}\rbrace \approx  \lbrace  \phi_{3},  \phi_{3}\rbrace \approx  \lbrace  \phi, \phi\rbrace \approx 0 .
\end{equation} 
Hence, we have the following set of equations
\begin{equation}
\left( \begin{array}{cccc}
0      & 0 &0 & \psi_{1}\\   
0  & 0 & 0  & \psi_{2} \\
0 & 0 & 0 & \psi_{3}\\
- \psi_{1}&-\psi_{2} &-\psi_{3} & 0\\
\end{array}\right)
\left( \begin{array}{c}
N     \\   
M  \\
Q \\
L\\
\end{array}\right)=0, \label{gff1}
\end{equation}
where $\psi_{1}$, $\psi_{2}$ and $\psi_{3}  $ are Poisson brackets of secondary constraints as
\begin{equation} 
\lbrace \phi_{1}, \phi_{2}\rbrace =0
,  \hspace{2mm}
\lbrace   \phi_{1},  \phi\rbrace =\psi_{1}
, \hspace{2mm}
\lbrace \phi_{2},  \phi\rbrace =\psi_{2}, \hspace{2mm}
\lbrace  \phi_{3},   \phi\rbrace =\psi_{3}, \hspace{2mm}
\lbrace  \phi_{2}, \phi_{3} \rbrace =0, \hspace{2mm}
\lbrace \phi_{1}, \phi_{3} \rbrace =0. \label{funcs}
\end{equation}
$\psi_{1}  $, $ \psi_{2} $ and $ \psi_{3} $ are actually similar to $\psi_{1}  $ and  $ \psi_{2} $
in Eq. (\ref{psi})for $ g $, $ f $ and $ l $ metrics respectively. 

Consistency equations (\ref{gff1}) may be satisfied in different ways for different sub-manifolds of phase space in which some of the functions introduced in Eq. (\ref{funcs}) vanish.
Similar to our analysis about Eqs. (\ref{gf}), we  see that if any one of the functions $\psi_{1}$, $\psi_{2}$ and $\psi_{3}$, are non-zero then at least one of the lapse functions should vanish, which is not acceptable. Hence the only physical possibility for consistency of secondary constraints is imposing new constraints 
\begin{eqnarray} &&
\psi_{1}\approx 0,\hspace{3mm}\psi_{2}\approx 0,\hspace{3mm}\psi_{3}\approx 0.\label{chosen} 
\end{eqnarray}
Inserting the constraints (\ref{chosen}) omits the last row and column of the matrix on the l.h.s. of Eq. (\ref{gff1}). Hence, the lapse function $L$ remains arbitrary which means the conjugate momentum $ P_L $ is a first class constraint.

Now, we should consider consistency of the chosen constraints (\ref{chosen}). This gives the following equations
\begin{eqnarray} &&
\left( \begin{array}{c}
\dot{\psi_{1}}     \\   
\dot{\psi_{2}}  \\
\dot{\psi_{3}}\\
\end{array}\right)=
\left( \begin{array}{ccc}
\mathcal{M}_{11}     & \mathcal{M}_{12} & \mathcal{M}_{13}\\   
\mathcal{M}_{21}     & \mathcal{M}_{22} & \mathcal{M}_{23} \\
\mathcal{M}_{31}    & \mathcal{M}_{32} & \mathcal{M}_{33}  \\
\end{array}\right)
\left( \begin{array}{c}
N     \\   
M  \\
‌Q\\
\end{array}\right)+\left( \begin{array}{c}
\mathcal{M}_{1L}     \\   
\mathcal{M}_{2L}  \\
\mathcal{M}_{3L}    \\
\end{array}\right)L=0,
\end{eqnarray}
where, $ \mathcal{M}_{11}=\{\psi_{1},\phi_{1}\} $, $ \mathcal{M}_{12}=\{\psi_{1},\phi_{2}\} $, $ \mathcal{M}_{13}=\{\psi_{1},\phi_{3}\} $,  $ \mathcal{M}_{21}=\{\psi_{2},\phi_{1}\} $, ...,  $ \mathcal{M}_{31}=\{\psi_{3},\phi_{1}\} $, $ \mathcal{M}_{1L}=\{\psi_{1},\phi\} $, 
..., $ \mathcal{M}_{3L}=\{\psi_{3},\phi\} $.
These are three equations for four unknowns $ N $, $ M $, $ ‌Q $ and $ L $. Considering $ L $ as the arbitrary lapse function, we obtain $ N $, $ M $ and $ Q $ in terms of $ L $ as follows
\begin{eqnarray} &&
\left( \begin{array}{c}
N     \\   
M  \\
Q\\
\end{array}\right)=-\left( \begin{array}{ccc}
\mathcal{M}_{11}     & \mathcal{M}_{12} & \mathcal{M}_{13}\\   
\mathcal{M}_{21}     & \mathcal{M}_{22} & \mathcal{M}_{23} \\
\mathcal{M}_{31}    & \mathcal{M}_{32} & \mathcal{M}_{33}  \\
\end{array}\right)^{-1}
\left( \begin{array}{c}
\mathcal{M}_{1L}     \\   
\mathcal{M}_{2L}  \\
\mathcal{M}_{3L}    \\
\end{array}\right)L.
\end{eqnarray}
One may change in advance the lapse functions to $\bar{N}  $, $ \bar{M} $ and $ \bar{Q} $ as follows
\begin{eqnarray} &&
\left( \begin{array}{c}
\bar{N}     \\   
\bar{M}  \\
\bar{Q}\\
\end{array}\right)=\left( \begin{array}{c}
N     \\   
M  \\
Q\\
\end{array}\right)+\left( \begin{array}{ccc}
\mathcal{M}_{11}     & \mathcal{M}_{12} & \mathcal{M}_{13}\\   
\mathcal{M}_{21}     & \mathcal{M}_{22} & \mathcal{M}_{23} \\
\mathcal{M}_{31}    & \mathcal{M}_{32} & \mathcal{M}_{33}  \\
\end{array}\right)^{-1}\left( \begin{array}{c}
\mathcal{M}_{1L}     \\   
\mathcal{M}_{2L}  \\
\mathcal{M}_{3L}    \\
\end{array}\right)L.\label{s0}
\end{eqnarray}
Introducing simultaneously, the new combination of second level constraints as 
\begin{eqnarray} &&
\phi^{\prime}=\phi-\left( \begin{array}{c}
\mathcal{M}_{1L} ,\  \mathcal{M}_{2L} ,\  \mathcal{M}_{3L}  \\
\end{array}\right)L\left( \begin{array}{ccc}
\mathcal{M}_{11}     & \mathcal{M}_{12} & \mathcal{M}_{13}\\   
\mathcal{M}_{21}     & \mathcal{M}_{22} & \mathcal{M}_{23} \\
\mathcal{M}_{31}    & \mathcal{M}_{32} & \mathcal{M}_{33}  \\
\end{array}\right)^{-1^{T}}\left( \begin{array}{c}
\phi_{1}   \\   
\phi_{2} \\
\phi_{3}\\
\end{array}\right), \label{s1}
\end{eqnarray}
the canonical Hamiltonian reads
\begin{eqnarray} &&
\mathcal{H}_{c}=\bar{N}\phi_{1}+\bar{M}\phi_{2}+\bar{Q}\phi_{3} +L\phi^{\prime}+L^{i}\mathcal{R}_{i}.\label{ka}
\end{eqnarray}
From Eqs. (\ref{chosen}), (\ref{s0}) and (\ref{s1}) it turns out that the constraints $\phi^{\prime}$  and the momentum $P_L$  would be two first class constraints. These two constraints together with $P_{L_i}$ and $\mathcal{R}_i$ constitute eight first class constraints which act as generators of space-time diffeomorphism. 

 Using the canonical Hamiltonian (\ref{ka}) it can be directly seen that consistency of $\phi_{2}, \phi_{1},\phi_{3}  $  as well as $ \psi_{1}, \psi_{2}, \psi_{3} $   would be satisfied if $ \bar{N}, \bar{M}, \bar{Q} $ are vanished. 
 In this way, besides the constraints $(P_{\bar{M}},P_{\bar{N}},P_{\bar{Q}})$ and  $(\phi_{2}, \phi_{1},\phi_{3})$ in the first and second level of consistency, we will have the constraints $(\psi_{1}, \psi_{2}, \psi_{3})$  and $(\bar{M}, \bar{N}, \bar{Q})$ in the third and fourth level. Hence, we found 12 second class constraints in this part of our analysis. Adding 18 second class constraints $P_{n_{(k)}^{i}}$ and $\mathcal{S}^{i}_{(k)} $, we have all together 30 second class constraints.

Counting the number of dynamical degrees of freedom then reads 
\begin{eqnarray} &&
\# DoF=80-16-30=34,
\end{eqnarray}
which corresponds to 17 degrees of freedom in configuration space concerning three massive and one massless gravitons. 

Note to the important role of the chosen constraints. In fact exactly in the subregion of the phase space specified by the constraints (\ref{chosen}) it is possible to find the required first class constraints to generate the gauge and simultaneously enough second class constraints to omit the Boulwer-Deser ghosts.

\subsection{Hamiltonian structure of multi-gravity} 
Adding up our experiences in three and four gravity, we can present a general Hamiltonian structure for multi-gravity. Consider a theory with $ \mathcal{N}-1 $  metrics $ g_{(k)\mu\nu} $ plus one metric $ h_{\mu\nu} $  where the Lagrangian consists of $ \mathcal{N} $ Hilbert-Einstein terms plus an interaction term of the form
\begin{eqnarray}&&
2m^{4}\sqrt{-g} \sum_{k=1}^{\mathcal{N}-1} \sum_{n=0}^{4} \beta^{(k)}_{n}e_{n}(\mathcal{K}^{(k)}), \label{a12}
\end{eqnarray}
in which $\beta^{(k)}_{n}$ are free parameters, $m$ is a mass parameter and the matrix $\mathcal{K}^{(k)}$ is $ \sqrt{(g_{(k)}^{-1}h)^{\mu}_{\nu}}$.
For this $ \mathcal{N} $-gravity model one should use the following  $ (\mathcal{N}-1)$ relations to change the shift variables as
\begin{equation}
N^{i}_{(k)}=L n^{i}_{(k)}+L^{i}+N_{(k)} D_{(k)j}^{i}n^{j}_{(k)}, \label{a115}
\end{equation}
where the $3\times3$ matrix $D^{i}_{(k) j}$ is similar to $(\ref{a18})$. Applying the above relations, the canonical Hamiltonian  linearizes  versus $ \mathcal{N}$ lapse variables as well as three shift variables $L_i$ as follows
\begin{equation}
\mathcal{H}_{c}=L \phi+N_{(k)} \phi_{(k)}+L^i \mathcal{R}_{i}. \label{hc}
\end{equation}
The momenta conjugate to  $N_{(k)}$, $n^{i}_{(k)}$, $L$ and $ L^{i} $ would be indicated respectively as   $P_{{(k)}}$, $P_{i(k)} $, $P$ and $P_{i}$ which are our primary constraints.
Consistency of these primary constraints leads to secondary constraints $\phi$, $ \phi^{(k)} $, $\mathcal{R}_{i}  $ and $ \mathcal{S}_{i(k)} $ where $\mathcal{R}_{i}  $ are defined similar to Eqs. (\ref{bii16}) and (\ref{bii6}) as $ \sum \mathcal{R}_{i}^{(k)} $  and $ \mathcal{S}_{i(k)} $ are resulted from the consistency of $ P_{i(k)} $ similar to Eq. (\ref{ti3}). Remembering the important property $\{\phi_{(k)},\phi_{(k)}\}\approx 0$ and 
assuming $ \{\phi_{(k)},\phi_{(k')}\}=0$  and $\psi_k=\{\phi,\phi_{(k)}\}$, consistency of the second level constraints $\phi , \phi_{(k)}$ gives
\begin{equation}
\left( \begin{array}{c|c}
& :\\
0  &  \psi_{(k)}\\
&  :\\
\hline   
..- \ \psi_{(k)} \ ..  & 0  \\
\end{array}\right)
\left( \begin{array}{c}
\colon     \\   
N_{(k)}\\
\colon\\
L\\
\end{array}\right)=0. \label{gff}
\end{equation}
Among so many different ways to satisfy the key equation (\ref{gff}), the only appropriate choice is considering $ \psi_{(k)}\approx 0 $  (see our discussion in conclusion section).
Furthermore, we should consider the consistency of the chosen constraints. This gives the following equations
\begin{eqnarray} &&
\left( \begin{array}{c}
:  \\   
\dot{\psi}_{(k^{\prime})}  \\
:\\
\end{array}\right)=
\mathcal{G}_{k'k}
\left( \begin{array}{c}
:     \\   
N_{(k)}  \\
:\\
\end{array}\right)+\left( \begin{array}{c}
:  \\   
\mathcal{G}_{k^{\prime}}  \\
:\\
\end{array}\right)L=0,
\end{eqnarray}
where $\mathcal{G}_{k'k} = \{\psi_{(k')},\phi_{(k)}\}$  and $\mathcal{G}_{k^{\prime}}=\{\psi_{(k')},\phi\}$. These are $\mathcal{N}-1  $ equations for $ \mathcal{N} $ unknowns $ N_{(k)} $ and $ L $. Considering $ L $ as the arbitrary lapse function, one can obtain  $ N_{(k)} $ in terms of $ L $ as follows
\begin{eqnarray} &&
N_{(k)}=-\mathcal{G}_{k'k}^{-1}\ \mathcal{G}_{k^{\prime}}
L.
\end{eqnarray}
One can change the lapse functions from the very beginning to $\bar{N}_{(k)}  $ where
\begin{eqnarray} &&
\bar{N}_{(k)} =N_{(k)}+\mathcal{G}_{k'k}^{-1}\ \mathcal{G}_{k^{\prime}} L, \label{z0}
\end{eqnarray}
and simultaneously introduce the new combination of second level constraints as 
\begin{eqnarray} &&
\phi^{\prime}=\phi-\mathcal{G}_{k^{\prime}}(\mathcal{G}^{-1^{T}})_{k'k}\ \  \phi_{(k)}L  .\label{z1}
\end{eqnarray}
Hence, the canonical Hamiltonian reads 
\begin{eqnarray} &&
\mathcal{H}_{c}=\bar{N}_{(k)}\phi_{(k)} +L\phi^{\prime}+L^{i} \mathcal{R}_{i}.\label{z3}
\end{eqnarray}
Now one can see that the constraints $ P_L $ and $ \phi^{\prime} $ are first class. Consistency of the second level constraints $ \phi_{(k)} $  as well as the chosen constraints $\psi_{(k)}$ at the third level would be satisfied if the modified lapse functions $ \bar{N}_{(k)} $ vanish altogether. Since, $ \bar{N}_{(k)} $ are conjugate to primary constraints $ \bar{P}_{(k)}$, we would have no further constraints due to determining the Lagrange multipliers via consistency of the last level constraints $ \bar{N}_{(k)} $. Our results are summarized in table 1.
\\

\hspace{15mm}\begin{tabular}{ |p{1.5cm}||p{2cm}| |p{2cm}| |p{2cm}||p{2cm}|}
	\hline
	\multicolumn{5}{|c|}{\textbf{Tabl 1 -Constraints Structure of Multi-gravity}} \\
	\hline
	Level& FC & FC &SC &SC \\
	\hline
	\vspace{2mm} 	level 1   & \vspace{2mm}  $ P $  & \vspace{2mm} $ P_{i} $  & \vspace{2mm}.... \vspace{2mm}$P_{i(k)} $\vspace{2mm}...&  \vspace{2mm}... $\bar{P}_{(k)} $ \vspace{2mm}...\\
	\hline
	\vspace{2mm} 	 level 2 &   	\vspace{2mm} $ \phi^{\prime} $ &  \vspace{2mm}$ \mathcal{R}_{i} $ &\vspace{2mm}... \vspace{2mm} $\mathcal{S}_{i(k)} $\vspace{2mm}... & \vspace{2mm}...\vspace{2mm} $ \phi_{(k)} $ \vspace{2mm}...\\
	\hline
	\vspace{2mm} 	 level 3 &   	      &&&\vspace{2mm}...\vspace{2mm} $ \psi_{(k)} $ \vspace{2mm}...\\
	\hline
	\vspace{2mm}  level 4 &   	   & & & \vspace{2mm}...\vspace{2mm} $ \bar{N}_{(k)} $ \vspace{2mm}... \\
	\hline
\end{tabular}
\vspace{6mm}

 Considering the Hamiltonian  (\ref{z3});  We have all together $ 8 $  first class constraints ($ P,P_{i},\mathcal{R}_{i}, \phi^{\prime} $) for generating the space-time diffeomorphism. We have also the set of
 $ (\mathcal{N}-1)\times 6  $ second class constraints $ (\mathcal{S}^{i}_{(k)},P_{i(k)}) $ in the forth columns 
 and   $ (\mathcal{N}-1) \times 4$ second class constraints in the fifth columns of table 1. 
   Adding up all of the contents of the table of constraints and using the master formula (\ref{mf}) we have
\begin{eqnarray} &&
{\color{black}\textbf{ $ \#DOF= 20\mathcal{N}-16-(\mathcal{N}-1) \times 4-(\mathcal{N}-1)\times 6 $ }}=2 \times (5\mathcal{N}-3),
\end{eqnarray} 
which correspond to $\mathcal{N}-1$ massive spin-2 and one massless spin-2 gravitons. 

As is seen our special choice of chosen constraints as $\psi_{(k)}$
provides enough additional constraints for omitting the ghosts as well as generating diffeomorphism. 

An interesting point is that the general pattern of constraint structure of multi-gravity happens also for bi-gravity, where at the last step, consistency of second level constraints in the lapse sector gives the following equations
\begin{equation}
\left( \begin{array}{cc}
0      & \Gamma \\   
-\Gamma  & 0  \\
\end{array}\right)
\left( \begin{array}{c}
N     \\   
M  \\
\end{array}\right)=0, \label{gff22}
\end{equation}
This should be compared with the general form of Eqs. (\ref{gff}) for multi-gravity. Our recipe for multi-gravity to assume $\psi_{(k)}$ as additional chosen constraints in this case reduces to consider just the function $\Gamma$ as the third level constraint. In fact, for bi-gravity we have not so many choices (non physical choice where both lapse functions $N$ and $M$ vanish). This is the reason why the bifurcation character of the problem is less seen for bi-gravity (see Ref. \cite{MS2} for a complete discussion).

\section{Conclusions }
Our main purpose in this paper was investigating the Hamiltonian structure of  four dimensional tri (multi)-gravity model in the context of ADM formalism. As in every other model of gravity the momentum conjugate to the lapse and shift functions are primary constraints. Based on the intellectual change of variables suggested by Ref. \cite{Hassan1}, (see Eq. \ref{a15}) the canonical Hamiltonian would be linear with respect to all lapse functions and one set of shift functions, say $L_i$ (This is, in fact, the main advantage of the interaction term between the metrics proposed in HR bi-gravity). Consistency of $P_i$ (conjugate to $L_i$) gives the sum of famous momentum constraints  which are first class and generate spatial diffeomorphism.  Consistency of the other momenta conjugate to other shift variables lead to a set of second class constraints. In this way there is no difficulty about the shift functions. 

However, consistency of the momenta conjugate to lapse functions is somehow challenging. Consistency of second level constraints in the sector of lapse variables gives a set of homogeneous linear equations for lapse functions (see Eqs. (\ref{gf}) for tri-gravity, (\ref{gff1}) for four-gravity and (\ref{gff}) for multi-gravity). This brings us to a bifurcation point, where some  lapse functions may vanish. This sounds nonsense physically, in fact one needs to consider a set of new constraints which decrease the rank of the matrix of coefficients, in order to avoid zero lapse functions.

	According to the theory, as far as one can deduce from the action by itself (i.e. with no subsidiary physical assumption)  one would be free to choose every set of  new constraints in addition to enough number of vanishing combinations of lapse functions such that the consistency equations of the second level constraints are satisfied. 
	
	However, detailed calculation of the consistency equations
	of the secondary constraint shows that if some of the functions  $ \psi_{(k)} $ are chosen to be non vanishing, then a number of lapse functions should vanish. 
	 Hence, the only physical choice is considering all functions $ \psi_{(k)} $ as new constraints.
	
Altogether, In order to have a physically acceptable ghost free theory for gravity, we should consider the following  criteria:

i) We need to find two  more first class constraints to complete the set of generators of the space-time diffeomorphism. 

ii) We need to  introduce enough additional second class constraints to omit ghosts.

iii) Non of the original lapse functions should vanish strongly. However, there may exist linear vanishing combinations of lapse functions. 

Fortunately, our analysis in this paper showed that the above criteria are satisfied on the constraint surface of new constraints $ \psi_{(k)} $.

 \end{document}